\documentclass[a4paper,11pt]{article}

\usepackage{jinstpub} 

\usepackage{fontawesome}

\title{\boldmath AlphaCAMM, a Micromegas-based camera for high-sensitivity screening of alpha surface contamination}

\author{Konrad~Altenm\"uller,}
\author{Juan~F.~Castel,}
\author{Susana~Cebri\'an,}
\author{Theopisti~Dafni,}
\author{David~D\'iez-Ib\'añez,}
\author{Javier~Gal\'an,}
\author{Javier~Galindo,}
\author[1]{Juan~Antonio~Garc\'ia\note{Corresponding author.}}
\author{, Igor~G.~Irastorza,}
\author{Gloria~Luz\'on,}
\author{Cristina~Margalejo,}
\author{Hector~Mirallas,}
\author{Luis~Obis,}
\author{Alfonso~Ortiz~de~Sol\'orzano}
\author{and Oscar~P\'erez}

\affiliation{Center for Astroparticles and High Energy Physics (CAPA), Universidad de Zaragoza, 50009 Zaragoza, Spain}

\emailAdd{juanangp@unizar.es}

\abstract{Surface contamination of \textsuperscript{222}Rn progeny from the \textsuperscript{238}U natural decay chain is one of the most difficult background contributions to measure in rare event searches experiments. In this work we propose AlphaCAMM, a gaseous chamber read with a segmented Micromegas, for the direct measurement of \textsuperscript{210}Pb surface contamination of flat samples. The detection concept exploits the readout capabilities of the Micromegas detectors for the reconstruction of \textsuperscript{210}Po alpha tracks to increase the signal-to-background ratio. We report here on the design and realization of a first 26$\times$26~cm$^2$ non-radiopure prototype, with which the detection concept is demonstrated by the use of a new algorithm for the reconstruction of alpha tracks. AlphaCAMM aims for minimum detectable \textsuperscript{210}Pb activities of $100$~nBq~cm$^{-2}$ and sensitivity upper limits about $60$~nBq~cm$^{-2}$ at 95\% of C.L., which requires an intrinsic background level of $5\times10^{-8}$~alphas~cm$^{-2}$ s$^{-1}$. We discuss here the prospects to reach these sensitivity goals with a radiopure AlphaCAMM prototype currently under construction. ~\href{https://github.com/rest-for-physics}{\faGithub}}

\keywords{Micropattern gaseous detectors MICROMEGAS, gaseous imaging and tracking detectors, particle identification methods. }

\arxivnumber{2201.01859}

\begin{document}
\maketitle
\flushbottom

\section{Introduction}
\label{sec:intro}

Rare event searches experiments aim at a background suppression down to negligible levels. In this context, a careful selection of radiopure materials close to the detector sensitive volume is required. A precise measurement of the radiopurity of the various components in the experimental set-up is performed using different techniques (see for instance \cite{exo1,next1,majorana,exo2,next2,xenon,gerda,lz}): gamma-ray spectrometry, Glow Discharge Mass Spectrometry (GDMS) and Inductively Coupled Plasma Mass Spectrometry (ICPMS).

The gamma-ray spectrometry technique \cite{gammaspec} uses semiconductor radiation detectors usually located underground. The gamma-ray emission of radioactive isotopes in the sample material can be detected, and the background contribution of the different materials can be assessed. High-Purity Germanium detectors (HPGe) are suitable for gamma-ray spectrometry because they offer very good energy resolution and low intrinsic background. This technique is non-destructive and is used to quantify the activity of different isotopes of radioactive chains. However, large samples of the target material together with large exposure times are generally needed in order to quantify low concentrations of radioactive isotopes. 
On the other hand, mass spectrometry techniques \cite{massspec}, such as GDMS and ICPMS, show the concentration of the different elements. However they cannot, in general, estimate the concentration of a particular isotope: for the natural chains of \textsuperscript{238}U and \textsuperscript{232}Th no information on daughters is typically obtained from mass spectrometry, unless secular equilibrium is assumed.

Although the techniques mentioned above have demonstrated to have a very good sensitivity to assess the radiopurity of the different materials, they are not adequate to measure \textsuperscript{210}Pb surface contamination in the target material.
Low background alpha spectrometers can be used to quantify bulk and surface contamination~\cite{Zuzel}; radiochemical methods to extract \textsuperscript{210}Po can be applied to increase sensitivity~\cite{Bunker}.
The lack of sensitivity of the gamma-ray and mass spectrometry techniques to \textsuperscript{210}Pb surface contamination has been spotted in the low-mass WIMP experiment TREX-DM \cite{trex-dm}. The radiopurity of all the materials close to TREX-DM sensitive area was measured using standard screening techniques, however, preliminary results point to an important contribution of \textsuperscript{210}Pb surface contamination to the TREX-DM background. The inconsistencies found between the TREX-DM background model and previously measured background levels exhibit the needs of a more precise measurement of surface contamination for the materials close to the active volume. In addition, \textsuperscript{210}Pb surface contamination on different materials has been shown to be a relevant background source also for other experiments searching for rare events like the direct detection of dark matter or the neutrino-less double beta decay (see for instance~\cite{gerda2,drift,lux,cuore,anais,yu}). The gamma-ray spectroscopy is not sensitive to alpha or beta emissions that do not have an associated high-energy gamma emission. Additionally, the GDMS and ICPMS techniques are destructive and require the manipulation of the samples, which itself can introduce further contamination.

\begin{figure}[htbp]
\centering 
\includegraphics[width=.8\textwidth]{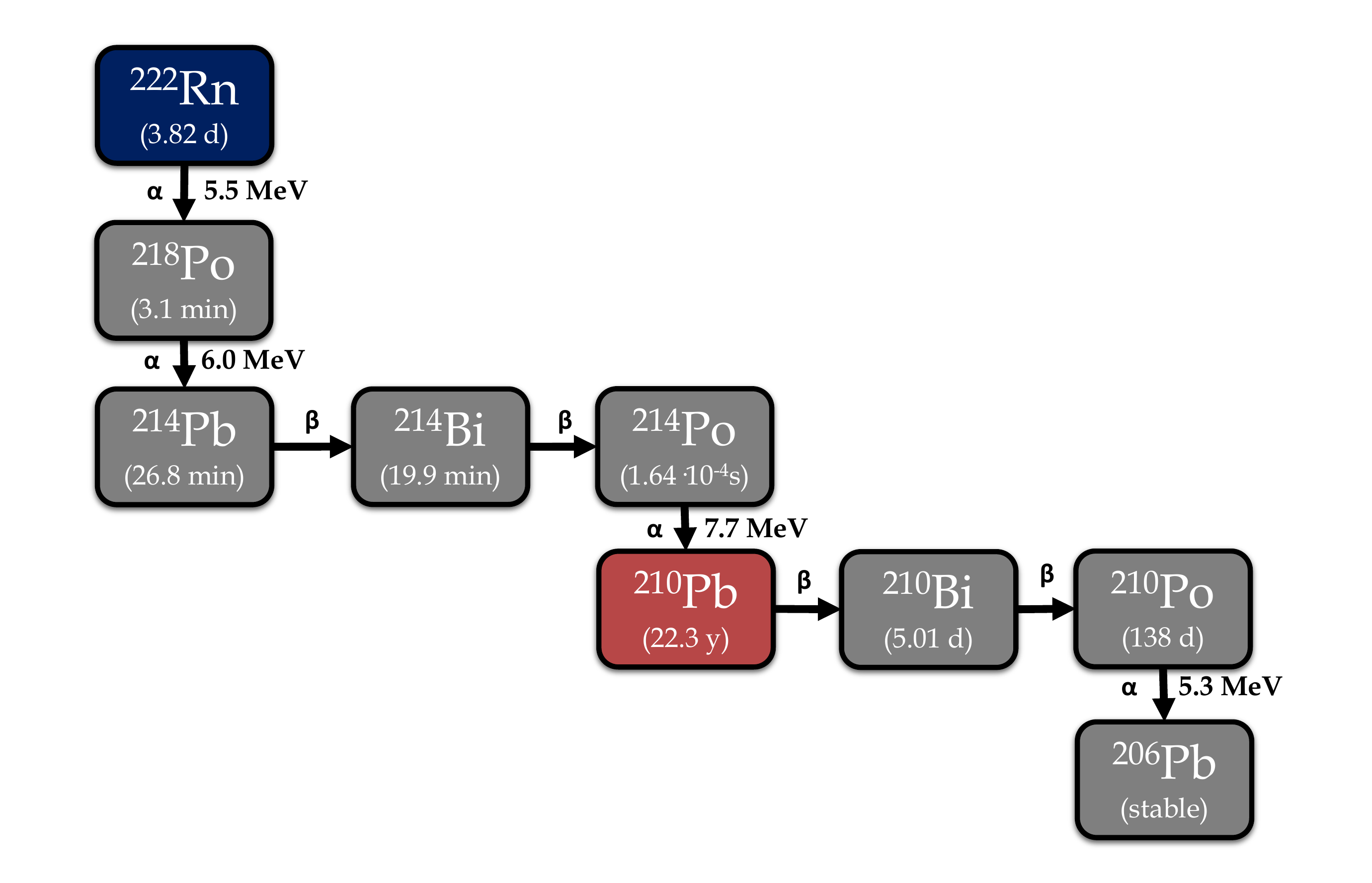}
\caption{\label{fig:RnDecay} \textsuperscript{222}Rn decay chain. The short-lived \textsuperscript{218}Po and \textsuperscript{214}Pb isotopes are positively charged after its decay and, therefore, will be electrostatically collected on negatively charged surfaces. In addition, this deposition could happen in other materials exposed to \textsuperscript{222}Rn during storage. These processes lead to the accumulation of the long-lived \textsuperscript{210}Pb isotope on material surfaces, which has been identified as a source of background for rare-event searches.}
\end{figure}

Exposure to \textsuperscript{222}Rn may lead to the contamination of material surfaces via the deposition of its progeny, e.g. during storage. Specific studies to quantify the rate of radon daughter plate-out and of different techniques to suppress it have been carried out for metals like copper and stainless steel and for other materials like polyethylene and polytetrafluoroethylene~\cite{snolab,Morrison,Guiseppe,Bruenner}. Moreover, \textsuperscript{222}Rn decay produces isotopes electrostatically charged, that could be attracted to the cathode inside an electric field~\cite{Pagelkopf}. \textsuperscript{222}Rn is emanated from surfaces with relatively high concentration of \textsuperscript{238}U and it tends to accumulate in non-ventilated areas. In that respect, one of the most dangerous isotopes is the \textsuperscript{210}Pb with a half-life of 22.3\, years (see figure~\ref{fig:RnDecay}) which tends to accumulate on surfaces exposed to \textsuperscript{222}Rn. The presence of \textsuperscript{210}Pb close to the sensitive volume of the detector is particularly dangerous: beta decays or characteristic X-ray emissions of the \textsuperscript{210}Pb and its progeny may lead to energy depositions in the region of interest for low energy searches (up to a few tens of keV).

Given the difficulties of measuring \textsuperscript{210}Pb surface contamination with standard radiopurity measurements, a novel detector design Alpha CAMera Micromegas (AlphaCAMM) is proposed in this paper. It will exploit the Micromegas readout capabilities to image and identify alpha particles directly stemming from \textsuperscript{210}Po alpha decays, which correspond to  \textsuperscript{210}Pb surface contamination assuming secular equilibrium (see figure~\ref{fig:RnDecay}). The motivation of AlphaCAMM is to complement the radiopurity measurement techniques described above in the context of TREX-DM or more generically in rare event searches experiments radio-assay programs. The main purpose is to measure samples with a very low concentration (<~100~mBq~kg$^{-1}$) of \textsuperscript{238}U and \textsuperscript{232}Th but with a potentially high out-of-equilibrium \textsuperscript{210}Pb surface contamination. Nonetheless, it should be noted that \textsuperscript{238}U and \textsuperscript{232}Th natural decay chains have several alpha emitters with decay energies similar to those of \textsuperscript{210}Po, a fact that complicates the  interpretation of the data.

The detection concept of AlphaCAMM together with its conceptual design will be introduced in section~\ref{sec:workingPrinciple}. In section~\ref{sec:discrimination}, a novel discrimination method based on the reconstruction of the direction of the alpha tracks will be described. The sensitivity prospects of AlphaCAMM will be discussed in section~\ref{sec:sens}.

During the development of this work, a similar concept has been proposed in~\cite{Screener} and~\cite{AIP}. In~\cite{Screener} a dedicated conceptual design and MonteCarlo simulations including the detector response are presented, while in~\cite{AIP} some experimental data are shown. In this work, an accurate event reconstruction of experimental data, together with sensitivity prospects, will be described.

\begin{figure}[htbp]
\centering
\includegraphics[width=.9\textwidth]{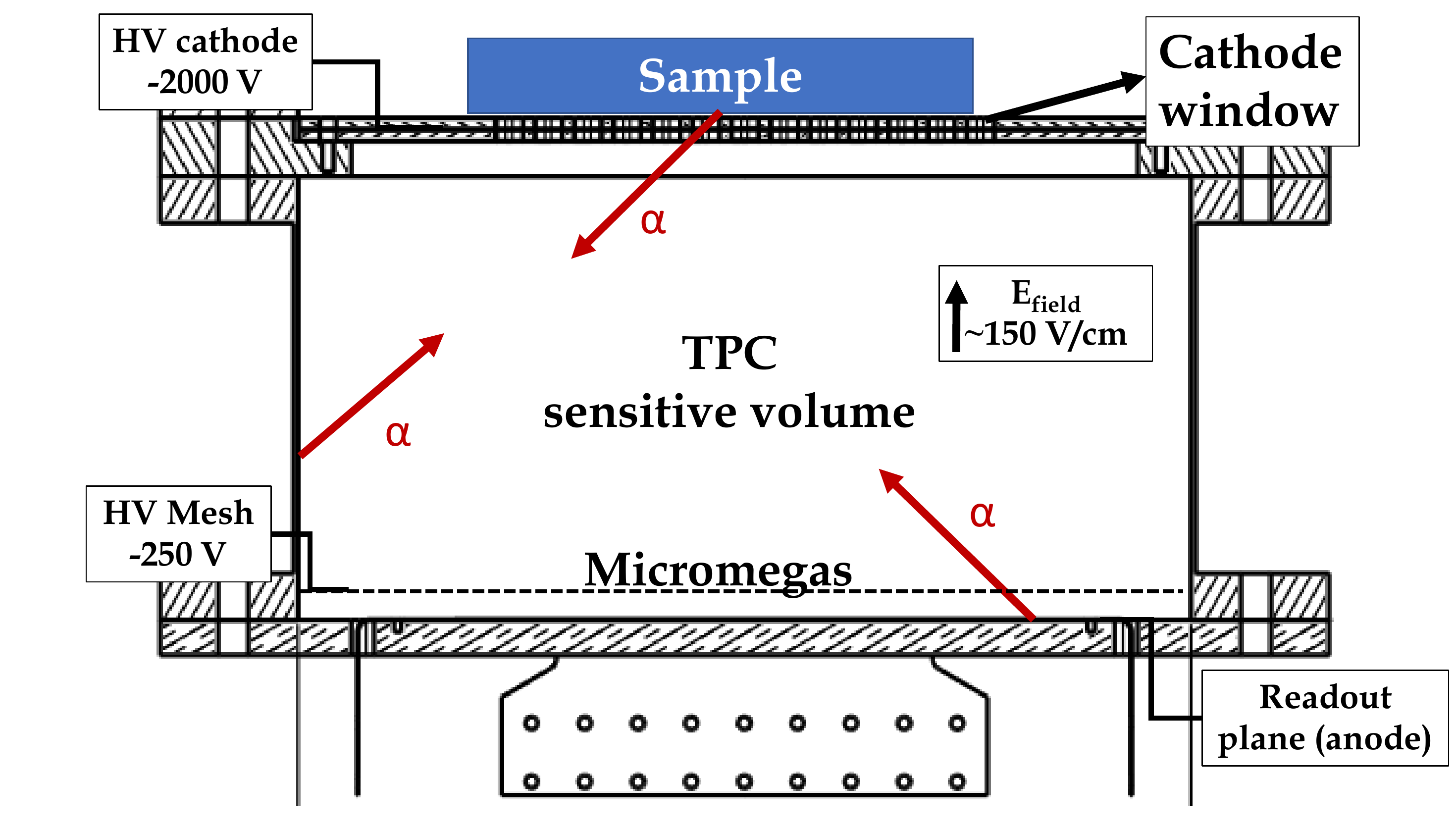}
\caption{\label{fig:workingPrinciple} AlphaCAMM working principle: The sample is located on the cathode window, which is transparent to alpha particles. The Micromegas readout plane provides topological information to reconstruct the origin and the end of alpha tracks. As shown, only alpha particles originated in the cathode and with a sense of movement towards the Micromegas readout contribute to the measurement.}
\end{figure}

\section{AlphaCAMM detection concept}\label{sec:workingPrinciple}

The detection concept of AlphaCAMM is based on the reconstruction of alpha particles emitted from a sample on a Time Projection Chamber (TPC). The sample is located on the TPC cathode, which is designed such as to minimize the energy loss of the crossing alpha particles and with a relatively high detection efficiency. The TPC is equipped with Micromegas (MICROMesh GAs Structure)~\cite{mm} readout planes. The latter comprises a metallic micro-mesh suspended over an anode plane made of strips or pixels, defining an amplification gap of the order of 50-100 {\textmu}m. Electrons drifting from the conversion region enter through the micromesh holes generating an avalanche inside the gap and inducing detectable signals. The Micromegas detectors offer good energy resolution and topological information of the events and, therefore, provide a powerful tool for the reconstruction of alpha tracks.

The working principle of AlphaCAMM is shown in figure~\ref{fig:workingPrinciple}. The sample to be measured is located on the cathode, which is made of a metallic grid with a relatively high transparency for alpha particles. The imaging capabilities of the Micromegas detectors allow the reconstruction of the direction of alpha tracks. Therefore, increasing the signal-to-background ratio under the assumption that only the alpha particles originated in the cathode contribute to the measurement.

\begin{figure}[htbp]
\centering 
\includegraphics[width=.9\textwidth]{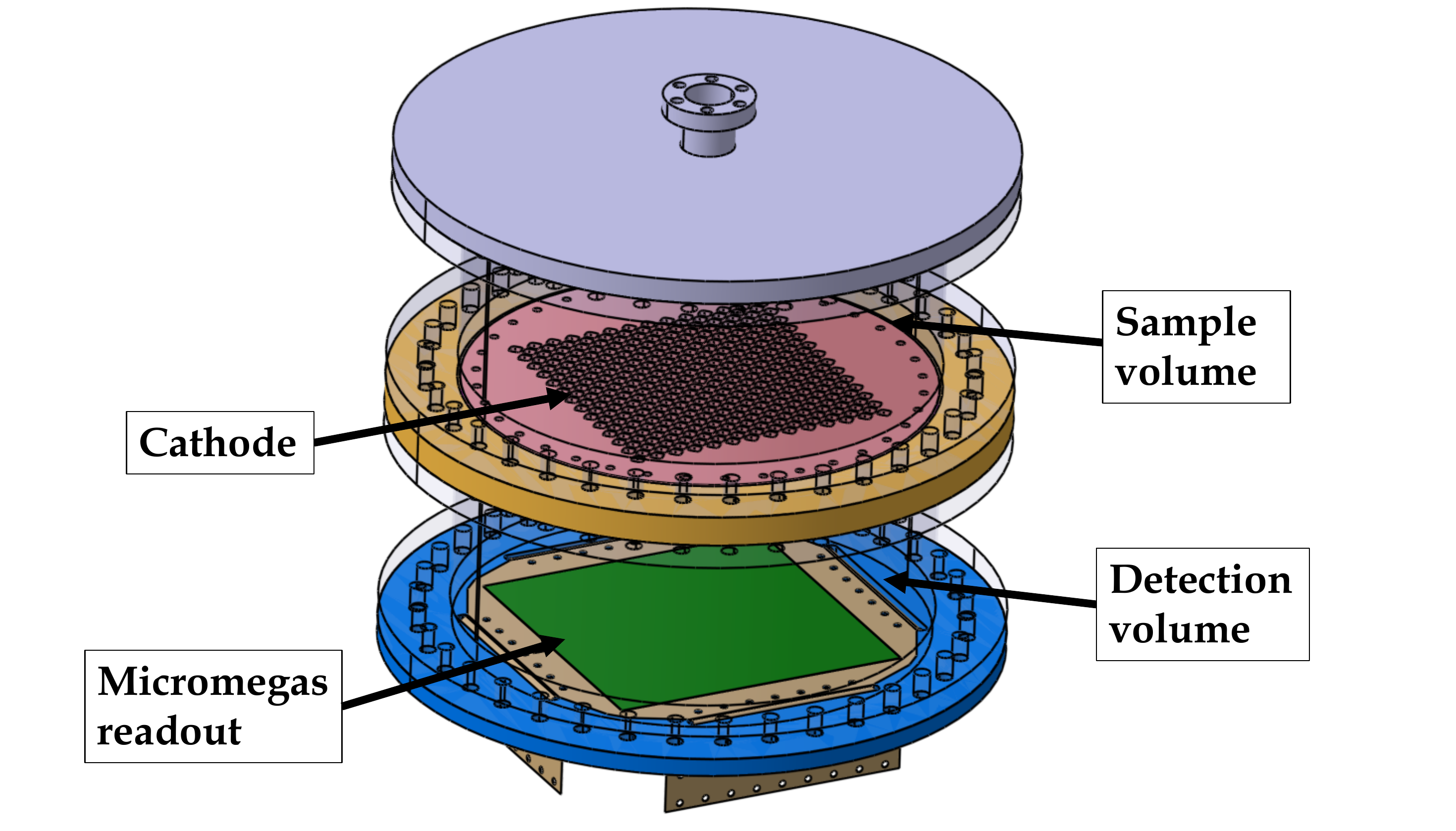}
\caption{\label{fig:alphaCAMMDesign} Conceptual design of AlphaCAMM. The chamber will be split into two different volumes: the sample will be placed at the upper part (sample volume), while the Micromegas readout plane will be located at the bottom (detection volume).}
\end{figure}

The design of AlphaCAMM is driven by the radiopurity requirements of the materials used in the construction of the chamber and the capabilities of the Micromegas readout planes for the reconstruction of alpha tracks. The chamber will be made of radiopure materials such as copper and polytetrafluoroethylene in an attempt to avoid possible \textsuperscript{222}Rn emanations from the innermost surface. The cathode grid will be made of radiopure copper with a design that ensures a high transparency to alpha particles. The preliminary design of the cathode implements a square grid pattern with a pitch of about 1~cm and a width of about 1~mm. An aluminized mylar foil of a few (2~-~5)~{\textmu}m thickness will be glued to the cathode grid, which corresponds to an energy loss from 0.4 to 1~MeV for alpha particles of about 5~MeV crossing it. The Micromegas will be manufactured with high standards of radiopurity using the microbulk~\cite{microbulk} technique, as shown in \cite{trexdm}. The readout plane will be made of strips with a pitch of $\sim$1~mm, that should provide a spatial resolution of about $500\,\mu$m for the reconstruction of alpha tracks. The readout will be performed using the AFTER~\cite{AFTER} electronics, which provides pulse digitization per readout channel and thus, time and energy information can be extracted with the granularity of single readout channels. The strip signals will be extracted via flat cables using a feedthrough at the bottom of the chamber to the electronics readout.

The conceptual design of AlphaCAMM is shown in figure~\ref{fig:alphaCAMMDesign}. The size of the chamber will be large enough to measure relatively large samples, for a first version of the detector we target a sensitive area of about 25$\times$25~cm\textsuperscript{2} and a height of 10~cm to allow the reconstruction of alpha tracks with energies up to 10~MeV. In order to guarantee a radon free environment, the chamber will be split into two different volumes separated by the cathode. The sample will be placed in the upper part of the chamber, while the Micromegas readout planes will be located at the bottom. The two separated volumes of the chamber should be leak-tight to be able to be pumped down so as to remove any pre-existing \textsuperscript{222}Rn contamination. The Micromegas detector will operate in an argon-isobutane mixture at a nominal pressure of 1~bar and in open loop. The upper part of the chamber will be flushed with nitrogen to avoid \textsuperscript{222}Rn in the sample space decaying and its progeny drifting and depositing on the top of the cathode, resulting in \textsuperscript{210}Pb contamination of the mylar. Alternatively, the upper part of the chamber will be connected with the bottom part and flushed with the argon-isobutane mixture used in the detection volume, therefore relaxing the mechanical requirements for the grid cathode pattern and the mylar thickness. In any case, the mylar cathode could be replaced regularly to avoid any possible \textsuperscript{210}Pb deposition.

\begin{figure}[htbp]
\centering 
\includegraphics[width=0.80\textwidth]{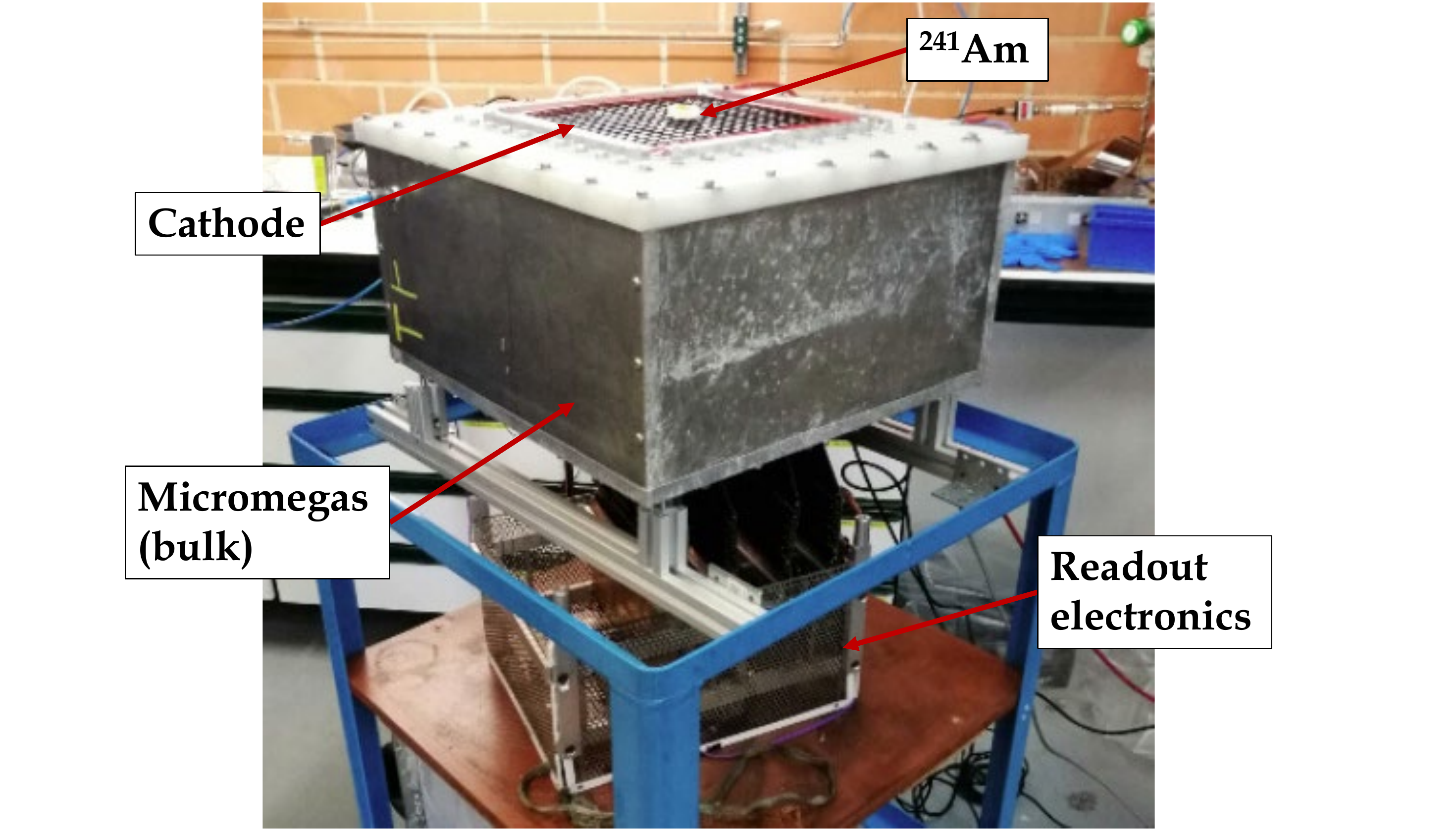}
\caption{\label{fig:prototype} Non-radiopure AlphaCAMM prototype, where all the main elements and the position of the \textsuperscript{241}Am source are shown.}
\end{figure}

\section{Signal discrimination}\label{sec:discrimination}

The signal discrimination will be performed using the readout capabilities of the Micromegas detectors. Alpha tracks follow a straight line with a large deposition of the energy towards the end of the track, also known as Bragg peak. This feature is used to identify the origin and the end of the track. In order to prove the performance of the Micromegas and the readout electronics to reconstruct the direction of alpha tracks, dedicated measurements in a non-radiopure AlphaCAMM prototype have been carried out. 

The experimental set-up is shown in figure~\ref{fig:prototype}, it consists of a bulk Micromegas with an active area of 26$\times$26~cm\textsuperscript{2} inside an aluminum chamber. The readout plane is made of strips with a pitch of 0.6~mm that results in 432 readout channels per axis. The reading of the strips is done using the AFTER electronics, which is triggered by the Micromegas mesh signal. The AFTER electronics are configured with a shaping time of 200~ns, a gain of 120~fC and a clock divider that provides a 40~ns-resolution per point. The measurements presented in this work have been carried out using an \textsuperscript{241}Am source, with an alpha particle energy of about 5.5~MeV, located at the center of the cathode.

\begin{figure}[htbp]
\centering 
\includegraphics[width=.95\textwidth]{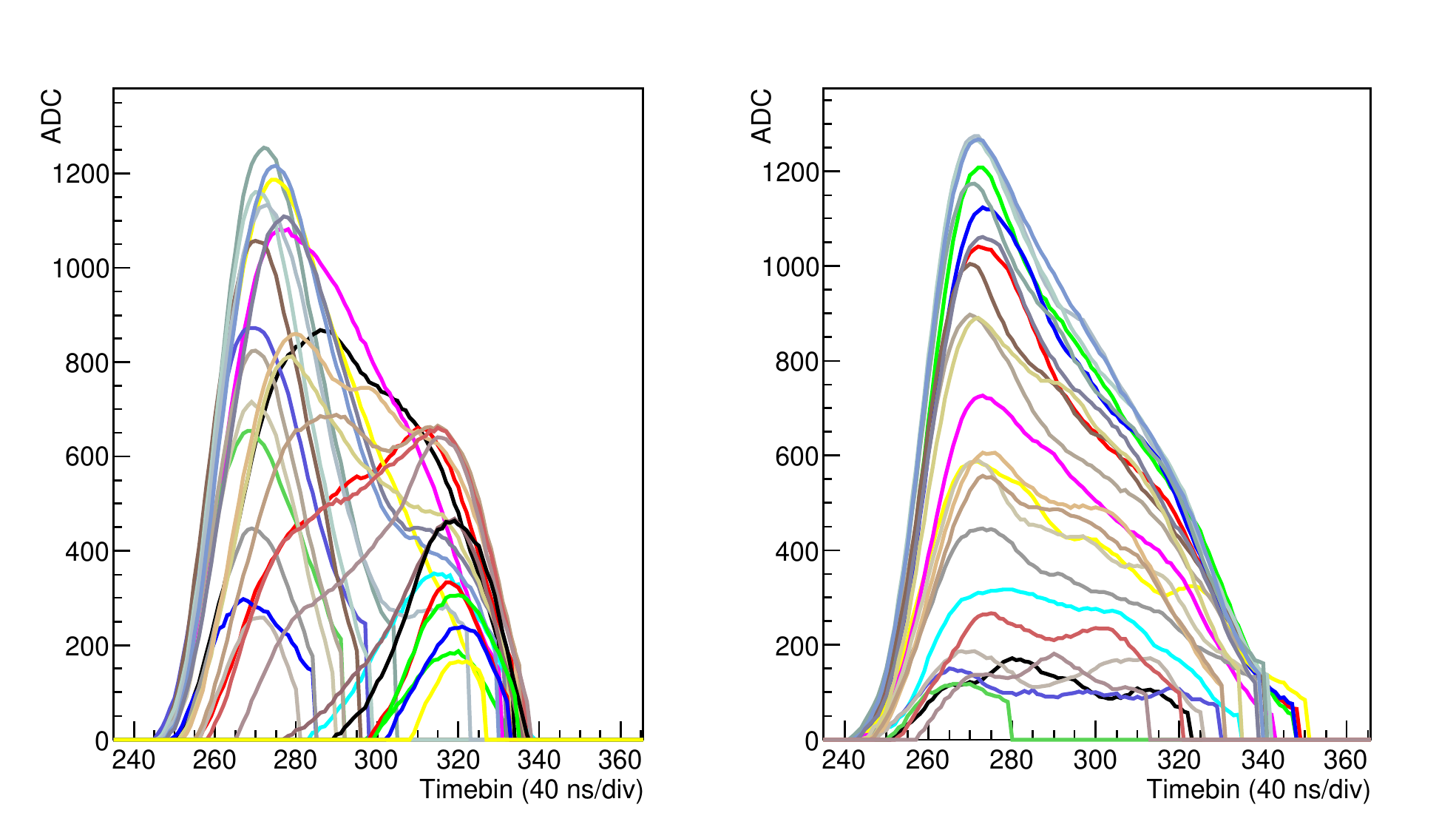}
\caption{\label{fig:calpulses} Digitized pulses from a single \textsuperscript{241}Am alpha event. The figure on the left shows the pulses for the X axis while the figure on the right shows the pulses for the Y axis. Only the pulses above threshold and after the baseline substraction are shown. The different colors represent different channels in the X and Y axis.}
\end{figure}

\begin{figure}[htbp]
\centering 
\includegraphics[width=.85\textwidth]{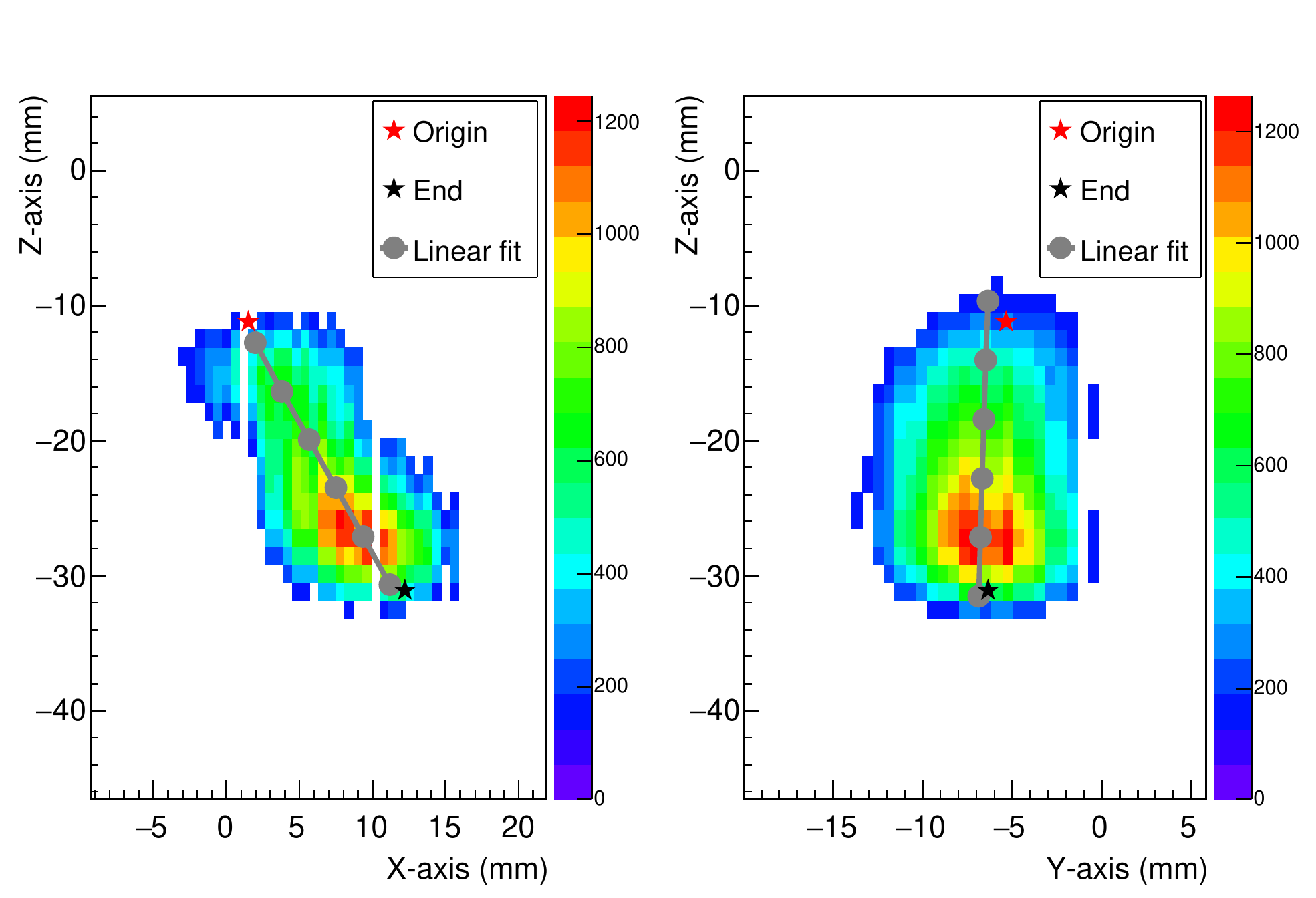}
\qquad
\includegraphics[width=.85\textwidth]{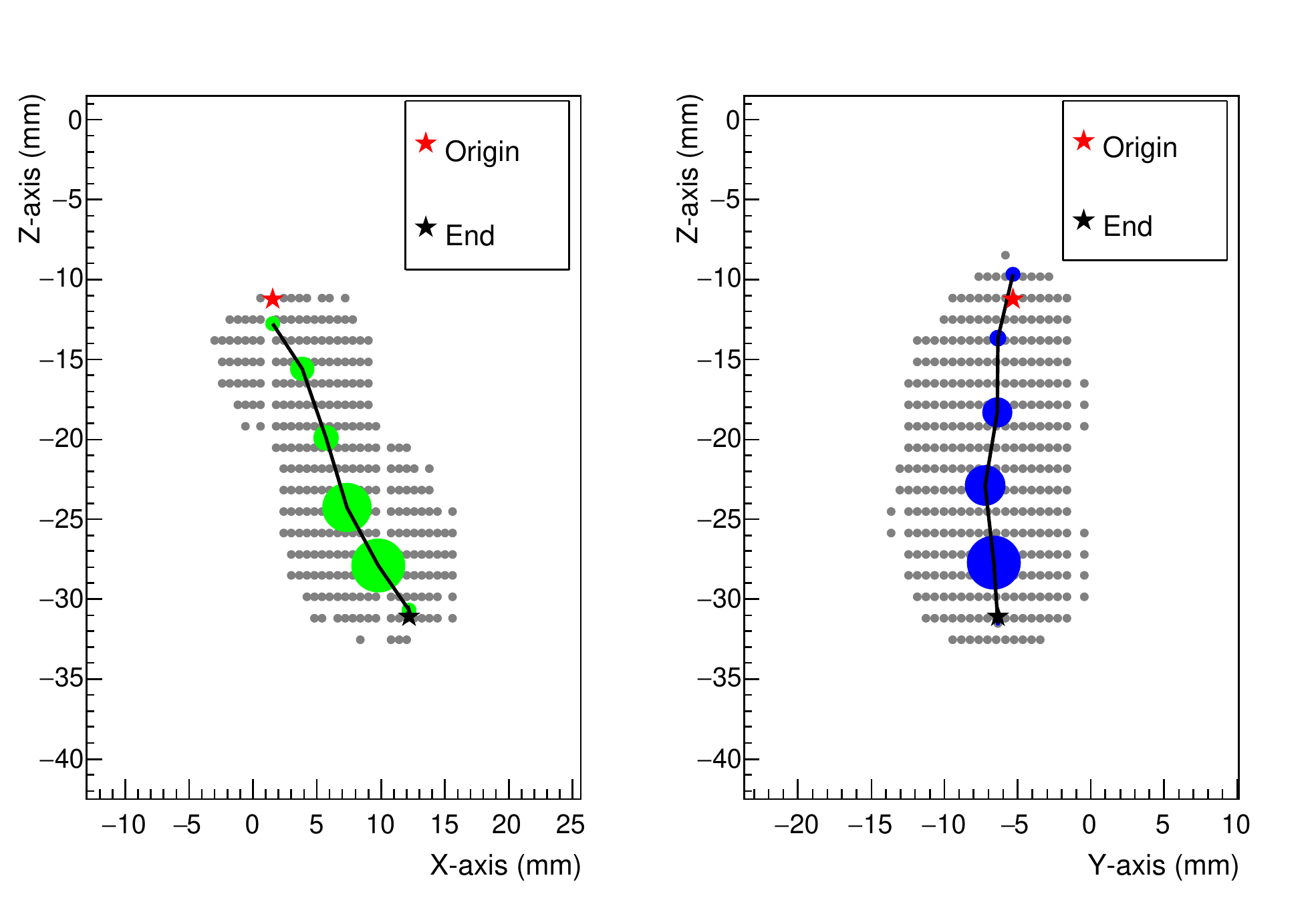}
\caption{\label{fig:calprojection} Top: Projection of a single \textsuperscript{241}Am alpha event pulses using an integrated bin size of 200~ns. The color palette shows the area of the pulses integrated during its corresponding timebin, in which the different \emph{hit} energy depositions are displayed. The figure on the left shows the XZ projection while the YZ projection is shown on the right. The grey lines represent the linear fit results in the different projections, while the grey dots are the selected nodes following the linear fit. The red stars represent the reconstructed origin of the track while the black stars represent the reconstructed end in the different projections. Bottom: XZ (left) and YZ (right) projections after performing the linearization of the alpha track, the green (left) and blue (right) circles represent the different energy depositions after performing the weighted average of the closest \emph{hit}, while the grey dots illustrate the different \emph{hits} shown on the figure at the top.}
\end{figure}

For the reconstruction of alpha tracks, the digitized pulses shown in figure~\ref{fig:calpulses} are integrated using a given timebin size in the XZ and YZ projections, which defines different \emph{hit} energy depositions. In this work, a bin size of 5 divisions, which corresponds to 200~ns, has been selected to match the shaping time used in the electronics. The different \emph{hit} projections in the XZ and YZ planes are shown in figure~\ref{fig:calprojection} top. Afterwards, two linear fits weighted by the energy deposition of the \emph{hits} are performed in the different XZ and YZ projections, the different linear fit results are  shown in figure~\ref{fig:calprojection} top. Following the linear fit results, different equidistant nodes are defined inside the region of the \emph{hit} depositions. In this work, six nodes have been selected to achieve a node spacing of about $5$~mm, and are displayed on figure~\ref{fig:calprojection} top. Later on, different energy depositions are obtained by computing the weighted average of the closest \emph{hits} to the nodes. Therefore, the track is reduced to six energy depositions close to the nodes defined after the linear fit. The different energy depositions after the track reduction are shown in figure~\ref{fig:calprojection} bottom. The track boundaries are extracted by using the most energetic deposition after the track reduction in the XZ and YZ projections. The origin of the track is determined as the further edge to the most energetic deposition, while the closest edge defines the track end. The Z coordinates which define the origin and the end of the track are extracted by performing the average value of the track boundaries of the XZ and YZ projections. The track is considered to be \emph{downward} when the difference between the origin and the end of the track in the Z direction is positive, while it is considered \emph{upward} otherwise. Preliminary simulations point out that a precise ($\sim500\,\mu$m) determination of the track boundaries in the X and Y axis is obtained using the reconstruction method described above.

\begin{figure}[htbp]
\centering 
\includegraphics[width=0.75\textwidth]{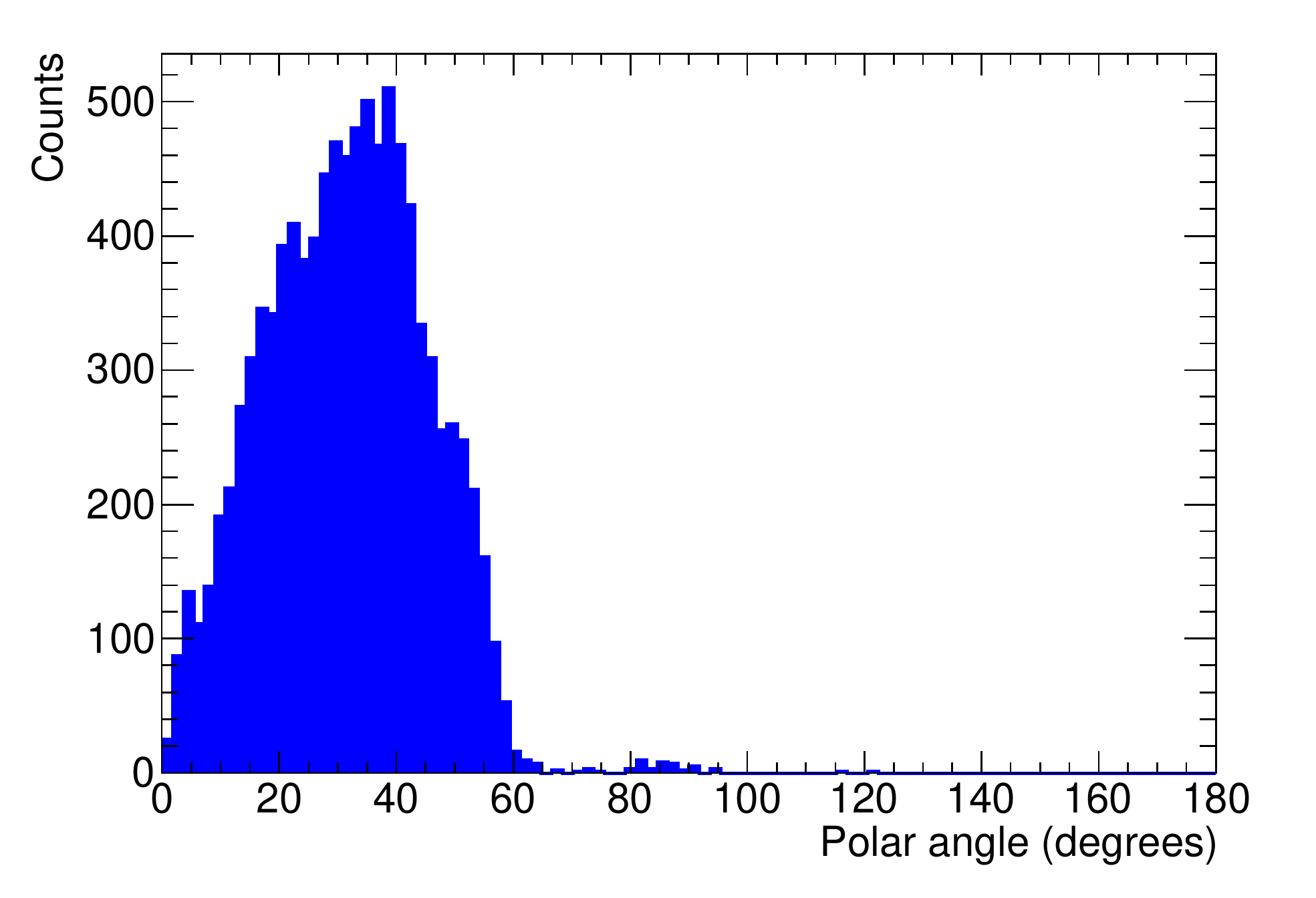}
\caption{\label{fig:angle} Reconstructed polar angle distribution of \textsuperscript{241}Am alpha tracks. Polar angles above 60° correspond to misidentified alpha tracks, which represent 0.5\% of the tracks. }
\end{figure}

\begin{figure}[htbp]
\centering 
\includegraphics[width=1.0\textwidth]{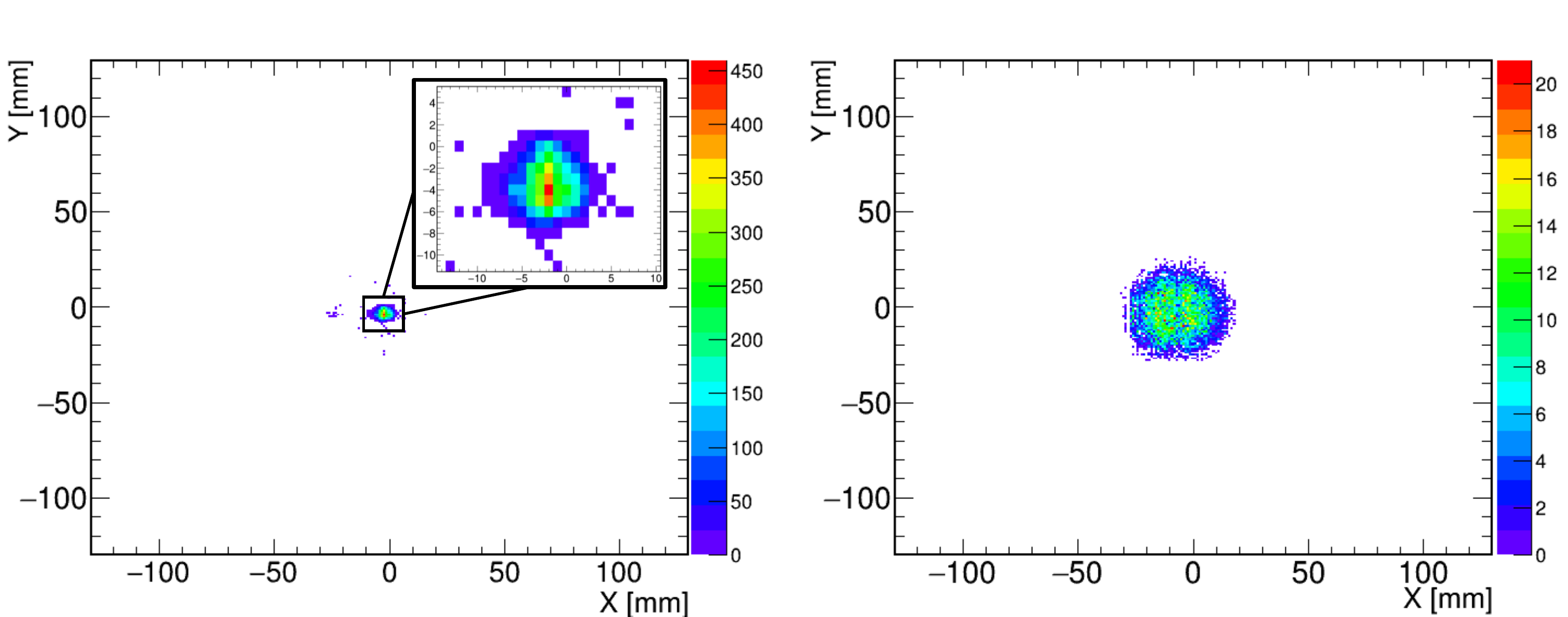}
\caption{\label{fig:originEnd} Left: Reconstructed origin of \textsuperscript{241}Am alpha tracks in the XY projection, where a zoomed view of the calibration hole is displayed. Right: Reconstructed end of the \textsuperscript{241}Am alpha tracks in the XY projection.}
\end{figure}

This reconstruction method has been tested with an \textsuperscript{241}Am alpha source located in the cathode of the chamber: 99.9\% of the tracks are properly identified as \emph{downward}. In addition, 99.5\% of the tracks are within a polar angle from 0 to 60°, as shown on figure~\ref{fig:angle}. The reconstructed origin and end of \textsuperscript{241}Am alpha tracks are shown in figure~\ref{fig:originEnd}. The calibration hole of about 5~mm of diameter is clearly seen in the reconstructed origin, while the isotropic direction of the track is displayed in the reconstructed end. The results presented before have been produced using REST-for-physics~\cite{rest}, a software framework for data analysis and simulations. New algorithms have been implemented at the track reconstruction library for physics events, used previously for topological reconstruction of electrons in neutrino experiments~\cite{PandaX}. The algorithms used in this work have been integrated at release v2.3.12~\cite{restRelease}. Additional track parameters such as the track length and the polar angle will be defined within the REST-for-physics framework and will be used for further discrimination.

\begin{figure}[htbp]
\centering 
\includegraphics[width=1.0\textwidth]{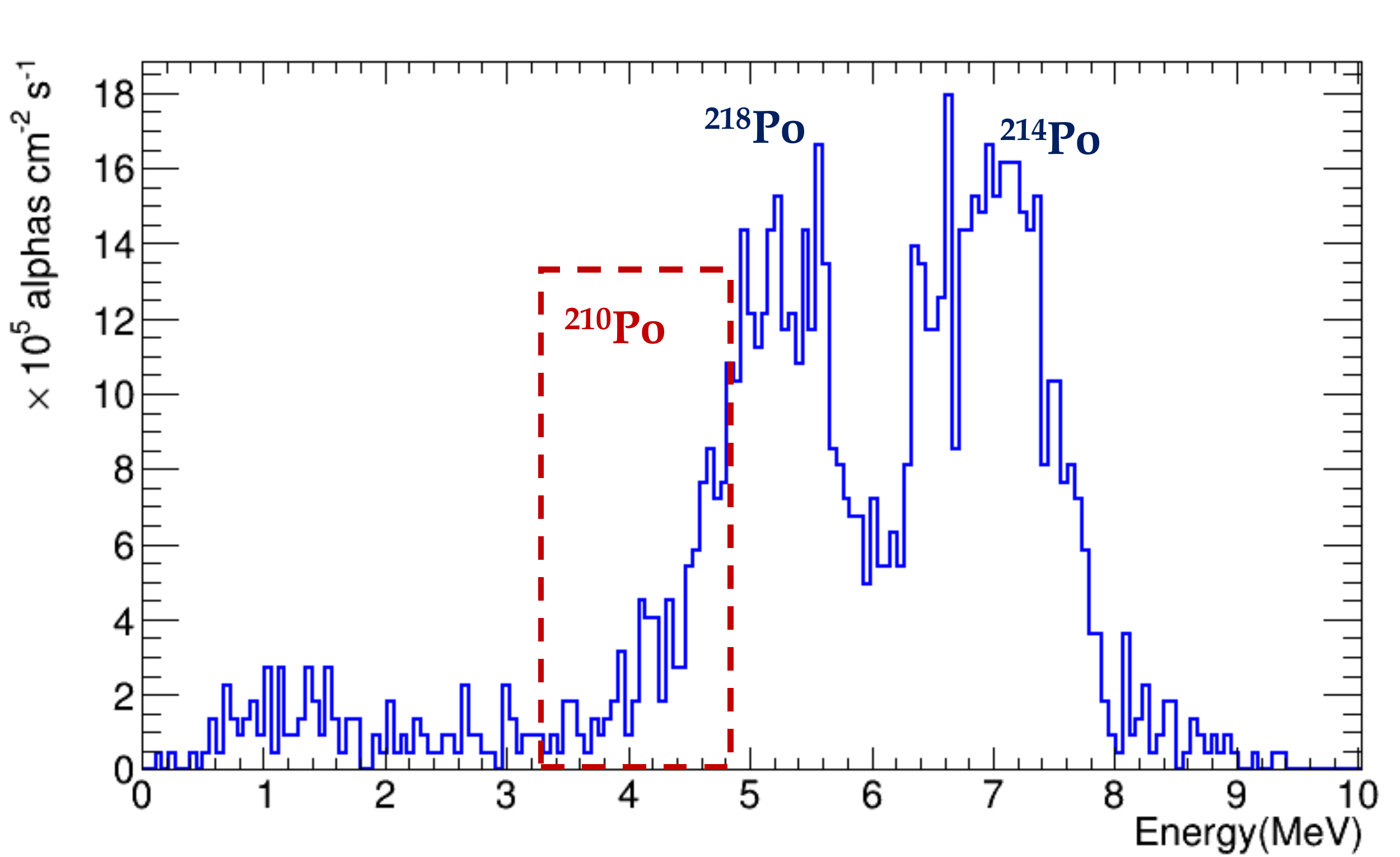}
\caption{\label{fig:spectra222RnProgeny} Background spectra of the non-radiopure AlphaCAMM prototype for alpha tracks with polar angles below 70°. Two different peaks corresponding to \textsuperscript{218}Po (6.0~MeV) and \textsuperscript{214}Po (7.7~MeV) alpha decays are visible. The red dashed area represents the expected position for the \textsuperscript{210}Po (5.3~MeV) alpha decay, which is not visible in the spectra. The energy resolution may be limited by the bulk Micromegas technology used in the prototype and gain fluctuations in the experimental set-up (due to its preliminary characterization in this first prototype). A further improvement in the energy resolution is expected by the use of microbulk Micromegas and a leak tight chamber proposed for the final design. Similar measurements were performed on~\cite{radondet}, obtaining much better energy resolutions.}
\end{figure}

Although AlphaCAMM will be built with high standards of radiopurity, non-negligible internal contamination cannot be ruled out. The reconstruction method presented before will allow the discrimination of the origin and the end of the tracks with an excellent precision. Under the assumption that \emph{downward} tracks are generated in the cathode while \emph{upward} tracks are generated in the anode, it will provide a powerful tool for background and signal discrimination. Moreover, the Micromegas readouts provide a good energy resolution that can be used to discriminate different isotopes from the \textsuperscript{222}Rn decay chain. As shown in figure~\ref{fig:RnDecay}, \textsuperscript{218}Po and \textsuperscript{214}Po decay via alpha emission with energies of 6.0~MeV and 7.7~MeV respectively, while the alpha decay energy of \textsuperscript{210}Po is 5.3~MeV. Some preliminary background data using the non-radiopure prototype described before are shown in figure~\ref{fig:spectra222RnProgeny}, in which two different peaks corresponding to \textsuperscript{218}Po and \textsuperscript{214}Po alpha decays can be distinguished. These measurements were performed with the cathode, made of an aluminized mylar foil, open to the air, that led to \textsuperscript{222}Rn progeny contamination of the cathode and justify the requirements of a \textsuperscript{222}Rn free environment for AlphaCAMM.

\section{Sensitivity prospects}\label{sec:sens}

The sensitivity of AlphaCAMM will be limited by its intrinsic background level. In a first stage, the background level of the chamber will be assessed using the reconstruction method described before, in which the different contributions from the cathode and the readout plane can be determined. One of the most dangerous contamination is the intrusion of \textsuperscript{222}Rn inside the chamber or the deposition of its progeny on the cathode. For this reason, the chamber has to be leak tight and should be pumped down or purged before the measurements; long exposures to the air have to be avoided. In addition, a radon free environment in the surrounding area of the cathode is required. The intrinsic background level of the chamber will be measured regularly and the mylar cathode will be replaced in case any dangerous contamination is detected. 

The \textsuperscript{210}Pb contamination of the sample will be determined by the number of measured alphas over the intrinsic alpha background of AlphaCAMM. Thanks to the imaging capabilities in the reconstruction of the alpha tracks, a geometrical cut in the area of the sample can be performed. The efficiency of the grid-patterned cathode for alpha particles $\epsilon$ will be calculated using simulations, that will be used to determine the surface contamination of the sample. Sensitivity prospects for AlphaCAMM have been estimated taking into account different background scenarios. Assuming a Poisson distribution of the data, the probability of obtaining $n$ counts for a mean of $\mu$ is given by:

\begin{equation}
\label{eq:p}
P (n|\mu) = \frac{\mu^{n} e^{-\mu}}{n!}\,.
\end{equation}

\begin{figure}[htbp]
\centering 
\includegraphics[width=0.8\textwidth]{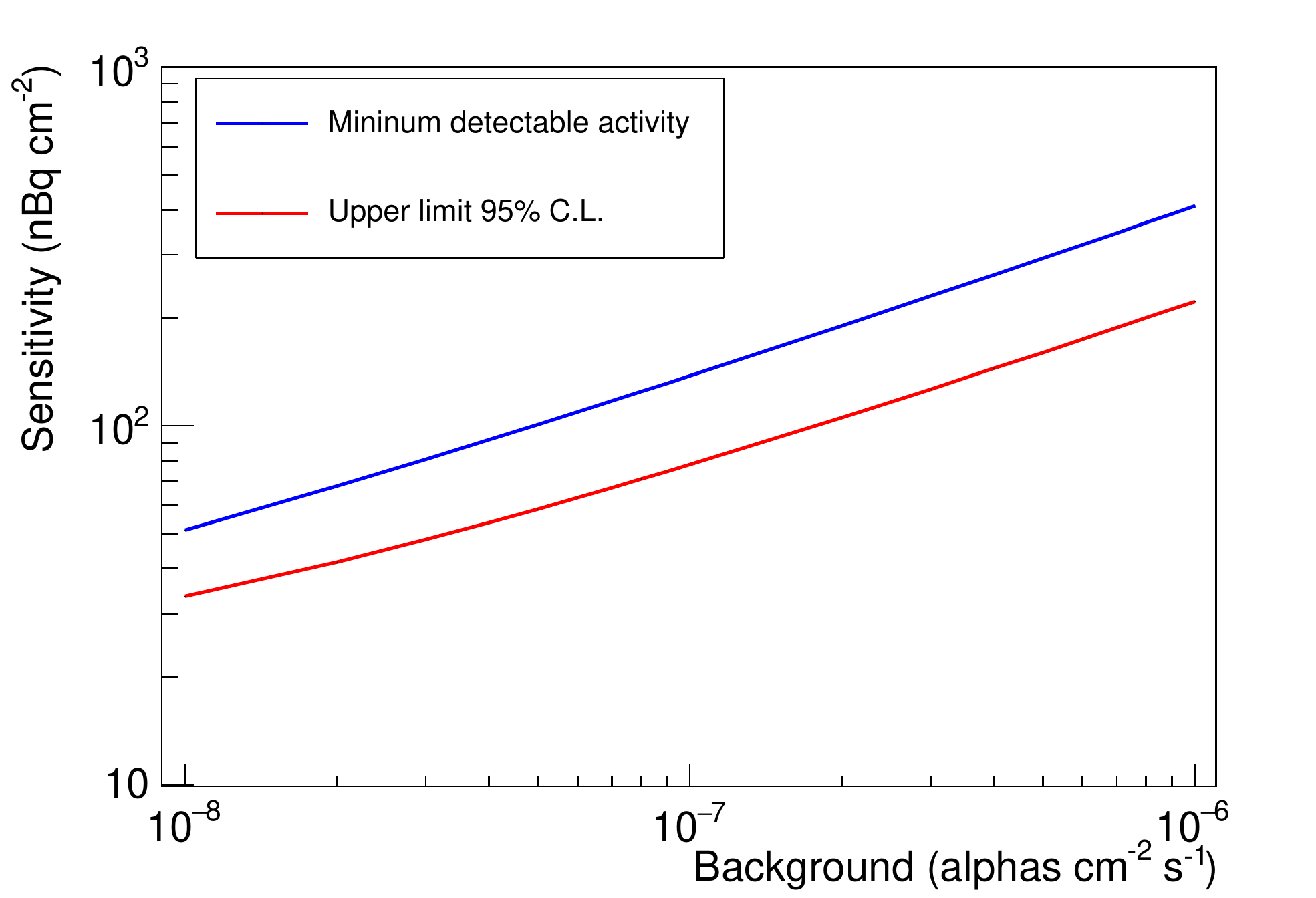}
\caption{\label{fig:sensitivity} Sensitivity prospects of AlphaCAMM versus its intrinsic background level. A time exposure of 1 week and a conservative efficiency of $\epsilon = 0.4$ for the grid pattern cathode have been assumed in the calculations. The blue line shows the minimum detectable \textsuperscript{210}Pb activity to reject the null hypothesis of \emph{background only}, with a significance of 3$\sigma$. The red line shows the upper limit for the \textsuperscript{210}Pb activity at 95\% of C.L. in case the null hypothesis of \emph{background only} cannot be rejected.}
\end{figure}

In order to determine the \textsuperscript{210}Pb activity of the sample, the null hypothesis of \emph{background only} has to be rejected. The $p$-value of obtaining $n$ events for a known background $b$ is given by:

\begin{equation}
\label{eq:gammab}
p (n,b) = \sum_{j=n}^{\infty} P (j|b)   = \frac{\gamma(n,b)}{\Gamma(n)}\,,
\end{equation}

\noindent where $\gamma(x,y)$ is the lower incomplete gamma function and $\Gamma(x)$ is the ordinary gamma function. Here we consider that a $p$-value of $0.001$, which corresponds to a significance of 3$\sigma$, is required to determine the \textsuperscript{210}Pb activity. In the case that the null hypothesis of \emph{background only} cannot be rejected, an upper limit at a 95\% of C.L. will be provided. The $p$-value of obtaining $n$ events for a known background $b$ and a signal $s$ is given by:

\begin{equation}
\label{eq:gammabs}
p (n,b,s) = \sum_{j=0}^{n} P (j|s+b)   = \frac{\Gamma(n+1,s+b)}{\Gamma(n+1)}\,,
\end{equation}

\noindent where in this case, $\Gamma(x,y)$ is the upper incomplete gamma function. Using equations~\ref{eq:gammab}~and~\ref{eq:gammabs} following~\cite{sens}, the sensitivity prospects of AlphaCAMM for different background levels have been estimated. The expected sensitivity of AlphaCAMM is shown in figure~\ref{fig:sensitivity}, in which background levels from $10^{-8}$ to $10^{-6}$ alphas~cm$^{-2}$ s$^{-1}$ have been explored. A sample of $625$~cm$^{2}$ with an exposure time of one week, which corresponds to $10.5$~h~m$^{2}$, and an efficiency of $\epsilon = 0.4$ for the grid pattern cathode have been assumed in the calculations.

The increase of the background in rare event searches experiments due to \textsuperscript{210}Pb surface contamination has to be carefully studied using a background model. In this work AlphaCAMM sensitivity goals are driven by the TREX-DM background model~\cite{trexdm}. Recent studies point out that \textsuperscript{210}Pb surface contaminations of about $100$~nBq~cm$^{-2}$ give a negligible contribution to the overall background of TREX-DM. Therefore, AlphaCAMM aims for minimum detectable activities of $100$~nBq~cm$^{-2}$ and sensitivity upper limits about $60$~nBq~cm$^{-2}$ at 95\% of C.L., that would require intrinsic background levels of about $5\times10^{-8}$~alphas~cm$^{-2}$~s$^{-1}$. This goal can only be achieved using high standards of radiopurity and ensuring a \textsuperscript{222}Rn free environment in the sensitive area and near the cathode.

The main contribution to the background level of AlphaCAMM is considered to stem from \textsuperscript{222}Rn contamination present in the argon mixture used as detection gas. This contribution may lead to activities of about $6.25~\mu$Bq in the sensitive volume of AlphaCAMM assuming \textsuperscript{222}Rn concentrations of about $1$~mBq~m$^{-3}$~\cite{Radon} in gaseous argon, even though this concentration could be diminished one month after production due to \textsuperscript{222}Rn decay. However, this activity cannot be easily extrapolated to the background level of AlphaCAMM. Assuming an isotropic emission of the alpha particles, up to 50\% of the alphas directly stemming from the \textsuperscript{222}Rn decay can be rejected using the reconstruction method presented in the previous section, therefore its contribution to AlphaCAMM background should be below $10^{-8}$~alphas~cm$^{-2}$~s$^{-1}$.

A further background contribution from the \textsuperscript{222}Rn progeny drifting to the cathode is expected, particularly the contribution of the short-lived isotopes \textsuperscript{218}Po and \textsuperscript{214}Po as shown in figure~\ref{fig:spectra222RnProgeny}. For the estimation of the \textsuperscript{218}Po and \textsuperscript{214}Po contamination, the upper part of the chamber has to be also taken into account because the positive electrostatically charged ions after the \textsuperscript{222}Rn decay in this region will be also attracted towards the cathode. Assuming the same \textsuperscript{222}Rn contamination and a similar volume in the upper and the lower part of the chamber, the \textsuperscript{218}Po and \textsuperscript{214}Po activities are estimated to be about $12.5~\mu$Bq in the cathode surface, that lead to activities of about~$20$~nBq~cm$^{-2}$. These activities can be extrapolated to background contributions of about $1\times10^{-8}$~alphas~cm$^{-2}$~s$^{-1}$ for the \textsuperscript{218}Po and the \textsuperscript{214}Po, under the assumption that up to 50\% of the alphas stemming from the \textsuperscript{218}Po and the \textsuperscript{214}Po are detected in the sensitive volume due to their isotropic emission, which lead to a total background contribution of $2\times10^{-8}$~alphas~cm$^{-2}$~s$^{-1}$. However, alpha particles directly stemming from \textsuperscript{218}Po and \textsuperscript{214}Po can be rejected due to their larger energy deposition. The contribution of \textsuperscript{214}Po, with an alpha particle energy of~$7.7$~MeV, to AlphaCAMM background can be considered negligible. On the other hand, the tail of the \textsuperscript{218}Po peak may fall in the energy range of \textsuperscript{210}Po as shown in figure~\ref{fig:spectra222RnProgeny}, in which 50\% of the \textsuperscript{218}Po are inside the red dashed area, where the \textsuperscript{210}Po is expected. In these conditions, the contribution of the \textsuperscript{218}Po to AlphaCAMM background is estimated at $5\times10^{-9}$~alphas~cm$^{-2}$~s$^{-1}$. Nevertheless, the energy resolution shown in figure~\ref{fig:spectra222RnProgeny} may be limited by the bulk Micromegas technology and possible gain fluctuations. As shown in~\cite{next}, energy resolutions below 2\% can be reached with a microbulk Micromegas detector operating in argon mixtures. In the case of those energy resolutions are reached in AlphaCAMM, the \textsuperscript{218}Po background contribution can be considered negligible.

The deposition of \textsuperscript{222}Rn progeny in the cathode may lead to \textsuperscript{210}Pb contamination, which is particularly dangerous for AlphaCAMM background. Nevertheless, the activity of the long-lived \textsuperscript{210}Pb isotope is considered to be low after the deposition of the \textsuperscript{222}Rn progeny, but it will require a careful monitoring of the background evolution of AlphaCAMM and a regular replacement of the mylar foil attached to the cathode. On the other hand, the contribution of \textsuperscript{222}Rn emanations from inside the chamber is considered to be negligible. \textsuperscript{222}Rn emanations from copper are estimated to be about $1$~$\mu$Bq~m$^{-2}$~\cite{Borexino}, which may led to activities of about $0.5~\mu$Bq in the chamber, a factor $12$ lower than the contribution assumed for the \textsuperscript{222}Rn concentration in the argon mixture. On the other hand, \textsuperscript{222}Rn emanations from the Micromegas readout planes of the microbulk type are considered to be negligible due to the high standards of radiopurity used in their construction~\cite{trexdm}.

Further contributions to the background level of AlphaCAMM are expected from the mylar foil attached to the cathode. The radiopurity of the mylar has been assessed in~\cite{trexdm}, obtaining upper limits of $0.59$~$\mu$Bq~cm$^{-2}$ for \textsuperscript{226}Ra. Although the mylar is intrinsically radiopure, most of the contamination occurs via surface deposition during the manufacturing process or during storage including the deposition of the \textsuperscript{222}Rn daughters. Therefore, dedicated measurements will be performed to asses the contribution of the mylar foil to AlphaCAMM background by measuring the intrinsic background level of the chamber with and without the mylar foil. In case of the mylar contribution is above AlphaCAMM requirements, the mylar foil can be eventually removed from the experimental set-up although it would require some modifications in the design presented before.

\section{Summary and outlook}\label{sec:summary}

Rare event searches experiments require a careful selection of radiopure materials close to the sensitive volume. However, standard techniques employed to assess the radiopurity of the materials are not sensitive enough to measure \textsuperscript{210}Pb surface contamination. A high-sensitivity detector based on Micromegas readout planes: AlphaCAMM, has been proposed to measure this contribution using alpha screening of \textsuperscript{210}Po. The motivation of AlphaCAMM is to complement other radiopurity measurements by quantifying \textsuperscript{210}Pb surface contamination in samples with a very low concentration of \textsuperscript{238}U and \textsuperscript{232}Th.

To demonstrate the detection concept of AlphaCAMM, a non-radiopure prototype has been developed, from which experimental data are shown in this paper. The readout capabilities of the Micromegas detectors have been proven by the development of a new reconstruction method for the determination of the origin and end of the alpha tracks with a high precision. AlphaCAMM will make use of this novel reconstruction method to discriminate if the alpha particle has been emitted from the cathode, where the sample will be located, or from the Micromegas readout plane and thus increasing the signal to background ratio. Further discrimination using the track energy and other track parameters such as the track length or the polar angle of the track will be studied. The excellent tracking capabilities of AlphaCAMM presented in this work have motivated a novel detector design with high standards of radiopurity, which is currently under construction.

AlphaCAMM sensitivity prospects in different background scenarios have been studied and it was shown that low alpha count levels are required to reach high sensitivities. Therefore it will be built with high standards of radiopurity. Moreover, the chamber has to be leak tight and a \textsuperscript{222}Rn free environment is required in order to avoid airbone  \textsuperscript{222}Rn intrusion and the deposition of its daughters in the cathode. The intrinsic background level of AlphaCAMM will be assesed once the chamber is built.


\acknowledgments

Authors acknowledge support from the the European Research Council (ERC) under the European Union’s Horizon 2020 research and innovation programme, grant agreement ERC-2017-AdG788781 (IAXO+), and from the Spanish Agencia Estatal de Investigaci\'on under grant PID2019-108122GB-C31. Authors also acknowledge the use of Servicio General de Apoyo a la Investigación-SAI, Universidad de Zaragoza. J.~A.~Garc\'ia acknowledges support from the Juan de la Cierva program from the Spanish Ministry of Economy and Competitiveness.


\end{document}